# Supernova Ia without Accelerated Expansion
# The First Global Failure of Relativity Theory

**Charles B. Leffert**   Wayne State University, Detroit, MI 48202, USA
E-mail: C_Leffert@wayne.edu

**Abstract.** A new cosmological model has been developed that shows great promise for solving many of the present problems of physics. A new concept of space and its production, "spatial condensation, SC," is the cause of the expansion. Dark mass (not matter) scales with the expansion differently than matter. Many other non-relativistic concepts predict a simple beginning, absence of singularities, a definition of energy and the cause of space curvature and gravity. Predicted cosmological parameters agree with recent measurements including $t_0$=13.5 Gy, $H_0$=68.6 km s$^{-1}$ Mpc$^{-1}$, $\Omega_{mass}$=0.28 and no dark energy. Other predictions include: Hubble flow at the Planck level, vacuum energy (no mass), $E_{vac}/E_{mass}=10^{123}$ in agreement with quantum mechanics, and the pattern in the CMB is the distribution of very early dark mass black holes of average mass $10^8$ $M_{sun}$. Excellent agreement with supernova Ia data is obtained with no acceleration of the expansion rate. It is concluded that the SC-model announces the first global failure of relativity theory.

**1. Introduction**

In the last decade of the twentieth century, astronomers having their new precision instruments, began measuring the luminosity of a number of exploding stars of a special class called: "supernova Ia (SNIa)" of relative constant maximum luminosity, which should be good cosmic distance markers. However, it was found that these data did not quite fit predictions of the current relativistic big bang (BB) model unless another term was added to their theoretical model. But this fit with a modified BB-model also predicted an acceleration of the expansion rate of our universe.

About the same time a completely new "spatial condensation (SC)" model for the expansion of our universe was developed [1]. On becoming aware of the BB fit with acceleration to the supernova Ia data, a check with this model was made. It produced an excellent fit to the supernova data **without any added dark energy** [2, 3] and **without acceleration** of the expansion. Its parameters are: the total average mass energy density, $\Omega_{mass} = 0.28$; age=13.5 Gy; Hubble constant, 68.6 km s$^{-1}$ Mpc$^{-1}$; and deceleration $q_0$=0.0084; but the new model had little resemblance to the current "lambda cold dark matter" (ΛCDM) model.

The success of this SC-model provided strong evidence that the SNIa measurements are revealing something much more profound than the need to add yet another unknown energy term to the already unknown and undetected mass energy called "dark matter." Instead, these first much more accurate measurements were implying the first global failure of relativity theory. The physics developed for one level of size of nature may fail at another. Relativity theory fails globally because it does not incorporate the cause that drives the expansion of our universe and it does not have sufficient spatial dimensions to account properly for that expansion.

Details of the new cosmological model have been presented elsewhere [1, 2, 3, 4, 5, 6, 7] but a summary is given below. The main aim is to further support the claim that no acceleration of the expansion is needed to account for the measured supernova Ia data.

There have been no experiments that directly challenged either the special or general theories of relativity theory or quantum field theory (QFT). However, the two major theories of general relativity (GR) and (QFT) are incompatible and there are problems understanding both. Since much effort has not mended the gap between these two theories, that failure suggests something very fundamental is missing in the foundation of physics. Even more serious, although seldom discussed, present physics has deficient concepts of what "*energy*" is [8], or "*time*" is, or "*space*" is. This paper, with its new dynamic at work in the universe, proposes a way to correct these deficiencies.

To convince the reader that the GR-BB-model is wrong in the absence of supporting experimental evidence, apart from the supernova Ia measurements, the proposed alternative model must predict the newly measured, high precision cosmological parameters with no adjustable parameters. Furthermore, one might hope such a model would point the way to solutions of other theoretical problems of present physics. The author will include frequent comparisons of detail to the current GR-BB theory.

**2. The New Vision of Our Universe**

**2.1 Morphology** Our universe is closed but it **CANNOT** collapse. This first sentence takes leave of relativity theory and the big bang model since both predict that a closed universe **CAN** collapse. Our spatially three-dimensional (3-D) universe is proposed to be the surface of an expanding four-dimensional (4-D) ball. The thickness of our 3-D universe in the 4-D direction is unknown but probably of order of the Planck length and much too small to be measured or to influence our 3-D gravity.

**2.2 Environment** The 4-D ball is expanding in an m-D "epi-space" that existed long before our 3-D universe and the 4-D ball came into being. All "*spaces*" are cellular and "*cosmic* time" is discrete. Unit cells of compacted 3-D and 4-D space are assumed to be incompressible cubes with Planck length edges and the m-D cells of epi-space are smaller yet.

Perhaps the most important new concept that should lead to the solution of many problems of present physics is that we live simultaneously in two universes with very different times and physics. Our closed, 3-D surface universe is limited by the non-Euclidean geometry of the expanding 4-D ball and its preferred reference frame. But the older epi-universe is not so limited, probably Euclidean, and can support processes and interactions with our 3-D universe that cannot be explained otherwise.

**2.3 Source of Expansion** The dynamic that is fundamental to the model is called "spatial condensation (SC)" and consists of the very small, higher dimensional cells of epi-space condensing to 4-D cells to cause the 4-D ball ,and thus our 3-D universe, to expand. This fundamental new concept of a "substantial space" is very different from the "infinitely stretching," non-substantial space of relativity theory. Also the concept of pervasive m-D cells of epi-space, in which we are contents, is very different than the current concept of compacted dimensional contents of 3-D space such as "strings."

**2.4 The Beginning** The very first 4-D cell produced was a foreign object, a catalytic site, in epi-space and triggered an exponential production of other (free) 4-D cells and the quick (~$10^{-33}$ s) compaction of the 4-D ball (4-D radius R~ 72 cm). The simple beginning is mathematically modeled elsewhere [1, 4]. Details of the end of formation of the 4-D ball set initial conditions for the eventual large-scale structure of galaxies and voids in 3-D space. One detail of the beginning of the expansion needed below is that two types of 4-D cells were produced: c-type acceptable to the 4-D ball and x-type rejected by the 4-D ball. C-type produce only c-type but x-type produce both c-type and x-type randomly but in the ratio of 3/1, respectively. The important x-type will become the dark mass of our 3-D universe. Radiation and matter came into being with the formation of the 4-D ball and, like the x-type cells, were rejected by the expanding 4-D ball.

**2.5 The Expansion** The production rate of 4-D cells is greatly reduced after the formation of the 4-D ball because the great excess of the (free) c-type 4-D cells are now compacted and shielded inside the 4-D ball. The 4-D ball itself now becomes the foreign object in epi-space with spatial condensation only on the outer surface layer of 4-D cells, and on the 3-D mass energy contents of radiation, matter and dark mass. (Geometrically, the radial expansion rate dR/dt of a spheroid of volume $V=k_1R^n$ must approach a constant C if the volumetric rate dV/dt is proportional to its surface $A=k_2R^{n-1}$ plus surface contents whose surface densities ρ(R) scale with R such that they decrease with the expansion.)

**2.6 Global Paradigm Shift** The fundamental consequences of the above radical change in vision of the beginning and expansion of our universe are very extensive. *Space* cannot be *stretched* as in the "inflation patch" to the BB-model, and the GR-Friedmann solution is rejected as a global model because the new gravity, without the attribute of attraction, cannot slow the SC-expansion of the 4-D ball and our 3-D universe.

However this change does not directly challenge the GR-Schwarzschild solution where local masses curve our 3-D space and bend light rays as verified for a number of the GR-predictions. Indeed, even though GR is handicapped by the missing fourth spatial dimension, it manages to invent its own "hyperspace" to predict at a point "a" outside a black hole, a radius R greater than the circumference through "a" divided by 2π [9]. This geometric "slight-of-hand' must be explained by the SC-fourth spatial dimension. To provide such understanding by the new SC-model, new physical concepts can be expected and relativity theory can be of help.

**2.7 Gravity and Quantum Behavior** This vision of the beginning and expansion of our 3-D universe is independent of any physics due to quantum behavior or gravity even though both of these phenomena are due to spatial condensation. These physical phenomena did not exist until the 4-D ball and our 3-D universe were born. Spatial condensation on clumps of mass does indeed curve our 3-D space and cause gravity, but only locally and without the attribute of attraction. Instead of slowing the expansion, an increase of mass in the universe increases the expansion rate.

**2.8 New Cosmic Time** The parametric time of present physics is symmetric but our subjective time of "past to present to future" is asymmetric. Fundamental physics has been concerned with local reversible phenomena over short time intervals where the expansion of our universe would have negligible effects. Here the expansion itself will be modeled, and spatial expansion is an irreversible process. There is no one single *time* to be found, so a new asymmetric cosmic time will be needed and the usual local laws of physics expressed in time-symmetric differential equations must be avoided without challenging their local use.

**3. Big Bang Spacetime**

The vision of Section 2 points in a new direction. The assumption that our 3-D universe is growing within a pre-existing greater universe of higher spatial dimensions immediately allows observers, in principle, to observe that expansion. With only three spatial dimensions, such observation is not allowed in relativity theory. With our universe interacting with, and within, the greater epi-universe means that our universe is not isolated and the first law of thermodynamics no longer applies globally in 3-D as stated in the third Friedmann equation of the BB-model but it may well apply in the total system of both universes.

However, these statements do not necessarily conflict with the well-established local laws of physics for short-time measurements. With our 3-D surface universe supported by a growing 4-D ball, the SC-model does not allow a collapsing universe as in the BB-model as a solution. Thus, in this new global model there is no possibility of building up a new cosmological model from either quantum theory or relativity theory. We start the expansion model from scratch and base it on simple new physical concepts. First, we will determine what can be deduced by taking seriously a cellular, substantial space.

**4. Consequences of Geometric Features**

In the following simple geometric SC-derivation, the contents of our 3-D space are ignored to discover what can be derived from just two geometric relations, Planck's natural units [$l_p$, $t_p$ and $m_p$] and the vision of Section 2. The key two geometric equations for the volumes of a 4-D ball $V_4$ and its surface $V_3$ are,

$$V_4 = \tfrac{1}{2} \pi^2 R^4, \quad \text{or } V_4 = \tfrac{1}{4} V_3 R, \tag{1}$$

$$V_3 = 2\pi^2 R^3. \tag{2}$$

Derivatives with respect to time give the expansion rates [the over-dot represents d/dt],

$$\dot{V}_4 = V_3 \dot{R} = 4 V_4 H, \tag{3}$$

$$\dot{V}_3 = 3 V_3 H, \quad \text{where } H = \dot{R}/R. \tag{4}$$

In terms of the of the number of compacted unit cells produced, $N_3$ and $N_4$ for 3-D and 4-D cells, respectively,

$$\dot{N}_4 = \dot{V}_4 / l_p^4 = 4 N_4 (\dot{R}/R) = 4 N_4 H \text{ and } H = 1/4\, \dot{N}_4/N_4, \tag{5a}$$

$$\dot{N}_4 = \dot{R}\, N_3/l_p = (N_3/t_p)(\dot{R}/c), \tag{5b}$$

$$\dot{N}_3 = \dot{V}_3/l_p^3 = 3N_3H \text{ and } N_3 = 4N_4(l_p/R). \tag{6}$$

Since in the outer layer, $N_{4L}=V_3l_p/l_p^4=N_3$, Eq. (5b) expresses Vision 2.3 of the SC-model well, where each of the 4-D spatial cells produce a new 4-D cell every Planck second, scaled for size by the factor $\dot{R}/c$. The 4-D ball responds by reproducing more 3-D spatial cells.

Vision 2.5 suggests the radial expansion rate of the spheroid, $\dot{R}$, will decrease and asymptotically approach a constant steady-state rate of $\dot{R}=C$. Here that constant is assumed to be the speed of light C=c and with increasing time R approaching R=ct. The recent measurements of the Wilkinson Microwave Anisotropy Probe (WMAP) [10] give the age of the universe as $t_0=13.7 \pm 0.2$ Gy ($ct_0=13.7 \times 10^9$ ly). If the radial expansion rate of this large universe has almost decreased to this limiting expansion rate, then the value of the radius is somewhat higher, say $14.2 \times 10^9$ ly or $R_0 \approx 1.35 \times 10^{28}$ cm. Accepting this estimate, then the Hubble constant $H_0 \approx c/R_0 = 68.9$ km s$^{-1}$ Mpc$^{-1}$. Thus the basic geometrical features of the model imply parameters in good agreement with recent astronomical measurements.

The kinematics of Eq. (5a) have accounted for the expansion of our 3-D universe by Eq. (5b) where the evolution with time is contained in the missing function $\dot{R}/c$ that must be obtained from other than geometric relations, but more can be deduced.

From Eq. (5b) for $\dot{R}/c=1$, the present total production rate of 4-D cells on the surface of the 4-D ball is $\dot{N}_{40} = N_{30}/t_p = 1.91 \times 10^{227}$ s$^{-1}$. The rate of spatial condensation $\dot{N}_4$ multiplied by Planck's constant $\hbar$ has units of energy. Therefore two new conjectures to be explored are the definition that *energy* of **all** types **is**:

$$E \equiv \dot{N}_4 \hbar, \text{ g cm}^2 \text{ s}^{-2}, \tag{7}$$

and that $E_{V0}=\dot{N}_{40}\hbar$ is the total vacuum energy in our present 3-D universe. The product $\dot{N}_3 \hbar$ also has units of energy but it is a secondary entity produced by the expansion of the 4-D ball. The unit 4-D cell, also called a "planckton," is a real entity that was produced in epi-space before our 3-D space came into being.

The units for energy were defined long ago in terms of mass, but these geometric consequences, so far, have made no mention of *mass*. Unlike Einstein's mass-energy $E=mc^2$, there is no factor of mass in Eq. (7), but like photon energy $E_v=\hbar v$, the unit of mass comes from Planck's constant. Using $\dot{N}_{40}$ above in Eq. (7) gives $E_{V0} \approx (N_3\hbar/t_p)(\dot{R}/c=1)$ $=2.01 \times 10^{200}$ ergs which is an enormous energy content for our universe with energy density approaching a fundamental constant value of $e_{V0}=E_0/V_0= \hbar/l_p^3 t_p=4.63 \times 10^{114}$ ergs/cm$^3$. The total mass energy density of our universe according to WMAP is $\Omega_{m0}=0.268$ and using $R_0=1.35 \times 10^{28}$ cm gives a total mass $M_{mu0}=1.15 \times 10^{56}$ g or a total energy due to mass of only $E_{m0}=M_{mu0}c^2=1.036 \times 10^{77}$ ergs.

Thus if Eq. (7) is to be acceptable for all energies, when applied to $\dot{N}_{40}$, $E_{V0}$ must be a new, different type of energy – one that does not carry the attribute of mass. Physics has already faced a similar dilemma with Casimir vacuum energy where theory had no need for mass to successfully predict measurements of the vacuum [11] so we tentatively call $E_{V0}$ the total vacuum energy, and $e_{V0}$ the vacuum energy density of the 3-D universe. Quantum mechanics also predicts a value for the vacuum energy of our universe and it is called: " the biggest and worst gap in our current understanding of the physical world" [12]. If wavelength is cut off at the Planck scale, consistent with the size of the 3-D spatial cells, quantum mechanics predicts the energy sum of ground state vacuum fluctuations is $10^{123}$ times the total mass energy content of our universe. This quantum prediction is easily checked with the above $E_{V0}/E_{m0} = 2.01 \times 10^{200}/1.03 \times 10^{77} \approx 10^{123}$, in agreement.

To check the consistency of the above use of Eq. (7) with mass energy, $\dot{N}_{4\,mu} = E_{m0}/\hbar = 0.981 \times 10^{104}$ planckton/s (pks) and dividing by the mass of the universe $\dot{N}_{4\,mu}/M_{mu} = 0.8522 \times 10^{48}$ pks g$^{-1}$ and multiplying by one Planck mass $m_p$ gives $1.855 \times 10^{43}$ pks and finally multiplying by one Planck time $t_p$ gives 1 planckton or one 4-D pk, sometimes called a "4-D hypercube." The last values can be derived directly from $\hbar^{-1}$ [2, App. 3].

As explained elsewhere [2, 4] for the SC-model, *mass, momentum, inertia* and *gravity* are due to the existence of **persistent** columns of arriving m-D spatial cells from epi-space. Other new SC-definitions are: *mass* **is** the number of such columns; *momentum*, **is** the angle $\theta_1$ of their arrival relative to **R**; *inertia*, **is** the resistance to change of $\theta_1$. Gravitational acceleration, $-GM/r^2$, occurs because the columnar impact of m-D cells from epi-space. It **is** equal to the 4-D epi-acceleration, $F_m/m = -GM/R_s^2$ times sin $\theta_2$ $= 4(l_p/m_p)^2(M/r)^2$ and $R_s$ is the Schwarzschild radius of a black hole of equivalent mass M. Thus sin $\theta_2$ is the local curvature of our 3-D space with no need for the attribute of "attraction" [2, 7]. All of the above have the attribute of mass connected to the attribute of persistent columns of m-D epi-cells to their massive condensation sites. The 4-D cells of outside layer of the surface of the 4-D ball are foreign objects in epi-space and support spatial condensation but in a different interrupted manner. Just as the incoming m-D cells from epi-space begin to form a persistent column on a 4-D cell, the deposit of another 4-D cell interrupts it and the process of column formation must begin again. Thus the vacuum mode of spatial condensation is different in that it occurs minus the attribute of "*mass*". In effect, the enormous "**Vacuum energy has no mass!**"

The fundamental concept of *action*, $\hbar$, has units of *energy* times *time*, which suggests that the important concept of "least action" translates to a frugal epi-space that tends toward "least spatial condensation," $N_4 \hbar = \dot{N}_4 \hbar \delta t$, and the product of one 4-D spatial cell in one Planck time **is** one unit of *action*.

Spatial condensation is uniform in our 3-D space on the surface of the 4-D ball and expands our universe everywhere. If the above Eq. (4), $\dot{V}_3 = 3V_3H$, is true globally, then it must also be true in a local volume $V_3 = (4\pi/3)r^3$. Using the Gauss theorem with a velocity v for the flow of new space out of an imaginary 2-sphere of radius r, then $\dot{V}_3 = 4\pi r^2 v$ and solving for v gives v=Hr, which is the equivalent of Hubble's law at the Planck scale. Note that, in principle, such a gravity-free (sin $\theta_2 = 0$) 3-D space would be a preferred reference system, where the divergence of v is uniform and any massive particle would have zero peculiar velocity to mark that position of substantial space. Such a particle in Hubble flow has no kinetic energy. It is at rest in the 3-D universe and can not impact another particle but other particles with peculiar velocities can collide with it.

Except for the new model of gravity, the above results flow from the geometry of a 4-D ball and the assumption of its steady-state expansion. Much has been derived from geometric features and a few visions, but more is needed to model the dynamics of the expansion contained in the function R(t) of Eqs (5a) and (5b).

**5. Consequences when Contents of 3-D Are Included**

How does one model the past evolution of our universe, from the birth of the 4-D ball to the present, so that it is consistent with the results already deduced from the above geometric assumptions? There are still the mass-energy contents of our universe that so far have been ignored but it was already shown that their contribution to the expansion was almost negligible by a factor of $10^{-123}$.

The BB-theory was developed with a mass energy component of non-interacting particles called " dark matter", the makeup of which is unknown and the presence of which has not been detected, except gravitationally. The dark mass to be built into the SC-model is neither particles nor matter but initial compacted collections of x-type 4-D cells that are rejected by the 4-D ball and should scale with the expansion differently than matter or radiation. The key variable of any cosmological model is the scale-factor function R(t), which in the SC-model is the radius of the 4-D ball and its 3-D surface. The past evolution of our universe is contained in the factor ($\dot{R}$(t)/R(t)) of relation (5a) and in factor $\dot{R}$(t) of relation (5b). So further progress requires development of R(t).

From the geometric consequences, it is the spatial condensation on the outer 4-D spatial cells of the 4-D ball (vacuum energy) that drives the expansion. Therefore the vacuum energy is also the indirect source of the changing mass-energy densities of the contents of the 3-D universe. Can the changing contents of the universe guide the evolution even though their present contributions to the expansion are so trivial, $10^{-123}$? The answer is yes; one must appeal to the changing contents of our 3-D universe and assume that their changing densities also affect the vacuum rate of spatial condensation.

It is encouraging to note that the early vacuum energy was much smaller on the surface of the beginning small 4-D ball (see Fig. 1) and future decreasing content densities would limit the vacuum energy density to a constant and the rate of expansion to the desired steady state. One must also get the correct definition of cosmic time.

With an appropriate definition of cosmic time, important scale factors for radiation and matter can be borrowed from the BB-theory for early nucleosynthesis. To obtain the limiting prediction ($\dot{R}$/c)$_\infty$=1 in the SC-contents model, it seems that a new and different scaling with the expansion is needed for the new third content of "dark mass".

**5.1 Contents Considerations** One observes that Planck units imply that: $G\rho_p t_p^2 = 1$ in which Planck time appears squared rather than linearly. This suggests that cosmic time t should also occur squared in a similar relation to the total density. Following the analogy that time acts like a resistance, cosmic *time* is defined as: "time **is** the resistance to spatial condensation." The three types of content vary in densities but act in parallel at the same time: so that like parallel electrical resistors they add by their inverse values.

To capture these thoughts, a concept of partial times $\Gamma_i$ was invented for each of the three contents (i=1,2,3) of our universe: radiation, matter, and dark mass, respectively, with $\Gamma_i^2 \equiv (\kappa/G)/\rho_i(R)$ where G is the gravitational constant and $\kappa$ is a new universal constant. In this non-relativistic expansion model, the "cosmic time t" of an event is defined as a function of the redshift Z of the universe but independent of the position, motion or redshift of the observer.

**5.2 Scaling of the Contents of the Universe with Expansion** The picture of spatial condensation and production of a substantial 3-D space has re-opened problems that are considered solved in the BB-model. The early BB-period of inflation of a stretching 3-D space is rejected and with it, so is the BB-notion of quantum production of very early fluctuations of mass densities that led to proto-stars and large-scale structure. Also re-opened are the horizon problem and the source of the homogeneity of our 3-D universe. An alternate early source of the seeds of small and large-scale structure in the 3-D universe must be found as well as new answers to the other problems..

The BB-model accounts for the scaling of density with the expansion of radiation as $\rho_r=\rho_{r0}(R_0/R)^4$ and baryon matter as $\rho_m=\rho_{m0}(R_0/R)^3$ where R is the scale factor and $R_0$ is its present value. The ratio can also be expressed in terms of the redshift Z as $R_0/R=(1+Z)$.

On the other hand, present physics has no physical evidence of the source or nature of the dominant mass called "dark matter" except that it does not radiate or act like normal baryonic matter. However, its density is still chosen to scale as normal matter

Temporarily call this unknown mass "x-stuff" or "dark mass" while deciding its scaling. The geometric consequences of the SC-model have already given a suggestion of the activity behind all types of energies including the new epi-energy of the vacuum that drives the expansion. It suggests that on large cosmic time scales, the epi-universe might decrease its rate of supply of epi-energy to one type of 3-D content, hold constant its rate of supply to a second content and even increase its rate of supply to a third content.

We can readily check this idea for the first two contents. We know that the volume of the universe increases as $V_u \propto R^3$ and Einstein's mass energy $E_u=M_uc^2$, so from the above scales, the photon radiation ($\rho_r \propto R^{-4}$) energy scales as $E_{ru}=\rho_r V_u c^2 \propto R^{-1}$ and the photons loose energy by an increase in their wave length during the expansion (cosmic energy is not conserved on large time scales.) The matter mass ($\rho_m \propto R^{-3}$) energy scales as $E_{mu}=\rho_m V_u c^2 \propto R^0 = 1$, and the epi-energy supply to matter tends to be conserved (apart from some local conversions to radiation)

This suggests that our needed seeds for condensation of matter might be the x-stuff mass with a scaling law of $\rho_x=\rho_{x0}(R_0/R)^2$ and even though its density decreases with the expansion, its mass energy increases as $E_{xu}=\rho_x V_u c^2 \propto R^{+1}$. But we gain much more. From Vision 2.4, with spatial condensation producing a new 4-D cell of x-stuff only on an existing cell of x-stuff, then this x-stuff grows as a clump of mass from the very beginning of our 3-D universe and there is no need for long-time gravitational instability to produce the seed as needed for the matter seeds of the BB-model. If the development is on the right track, the model is expected to produce its own explanation for its homogeneity.

With the scaling of dark mass decided, attention is returned to the development of a cosmic time for the model. The partial time for dark mass is $\Gamma_x^2=(\kappa/G)/\rho_x(R)$ where the constant $\kappa$ is not yet set. Continuing with the resistance analogy for time, the three partial times act in parallel, so the decision is made to sum their parallel contributions to the expansion by their inverse squares (parallel electrical resistors sum inversely):

$$t^{-2} \equiv \Sigma_i \Gamma_i^{-2}. \tag{8}$$

**5.3 Completion of the SC-Model** Equation (8) will be used again to connect the cosmic times before and after formation of the 4-D ball. The universal constant $\kappa$ ($= G\rho_T t^2$), of fundamental value, canceled out of the final expression for the desired asymmetric time,

$$t = +(t_0^2 \, \rho_{T0}/\rho_T(R))^{1/2} \, , \tag{9}$$

$$\kappa = Gt^2\rho_T = Gt_0^2\rho_{T0} = 3/32\pi, \tag{10}$$

where $\rho_T = \rho_r + \rho_m + \rho_x$. The value for the universal constant $\kappa$ was set to $3/32\pi$ the same as for radiation alone in the BB-theory.

We have yet to derive the cosmological parameters and their predicted values from these relations; and to check whether those predictions for the present agree with the SC-geometric consequences; but the complete expansion model is fully stated in Eqs. (9) or (10). Thus, in this SC-model, cosmic time is stated directly as an explicit t(R) relation rather than as a differential equation for R(t).

The time derivative of $\rho_T$ is $\dot{\rho}_T = -2H\rho_{T2}$, where $\rho_{T2} = 2\rho_r + 3/2\rho_m + \rho_x$. Solving Eq. (10) for $t^2$ and taking the time derivative gives,

$$tH = \rho_T/\rho_{T2}, \tag{11}$$

where $H = \dot{R}/R$. It is clear that the range is $½ \leq tH \leq 1$ from early dominance of radiation to late dominance of dark mass, x. In Eq. (11) write H as $\dot{R}/R$ and introduce the speed of light c on both sides of the relation to obtain,

$$\dot{R}/c = (R/ct)(\rho_T/\rho_{T2}), \tag{12a}$$

or,

$$\frac{\dot{R}}{c} = \left(\frac{R_0}{ct_0(1+Z)\rho_{T0}^{1/2}}\right) \frac{[\rho_{r0}(1+Z)^4 + \rho_{m0}(1+Z)^3 + \rho_{x0}(1+Z)^2]^{3/2}}{[2\rho_{r0}(1+Z)^4 + 3/2\,\rho_{m0}(1+Z)^3 + \rho_{x0}(1+Z)^2]}. \tag{12b}$$

The range is $½(R/ct)_{eoc} \geq \dot{R}/c \geq 1$. It is clear from Eq.(12b) that this dynamical model has the same steady-state limit $\dot{R} = c$ as the SC-geometric consequences. Note integer and integer fraction limits.

These and other relations were incorporated into a computer program with an input variable of $R/R_0 = 1/(1+Z)$ that could be varied from the first 4-D cell produced to far into the future universe. Program input constants included: present radiation density $\rho_{r0}$ = 9.40x10$^{-34}$ g cm$^{-3}$, fixed by the present CMB temperature of 2.736 K; baryon density $\rho_{m0}$ = 2.72x10$^{-31}$ g cm$^{-3}$, fixed by a fit to early nucleosynthesis calculations; [13, 14]. The one remaining program constant was the age of the universe $t_0$, however that constant has now been fixed by WMAP to 13.7 ± 0.2 Gy. Therefore the SC-contents model of the expansion is complete. Next, we explore further predictions and whether those predictions agree with the SC-geometric model.

### 6. Predictions of the SC-Model

**6.1 Agreements** The total average density $\rho_{T0}$ of the mass contents of the universe is fixed by Eq. (10) for the universal constant $\kappa$. The current average density of dark mass is set by the balance, $\rho_{x0} = \rho_{T0} - \rho_{r0} - \rho_{m0} = 2.19 \times 10^{-30}$ g cm$^{-3}$ giving a ratio of $\rho_{x0}/\rho_{m0} = 8.05$ in good agreement with astronomical measurements for both baryon density (here, $\rho_{m0}$) and mass density (here, $\rho_{m0}+\rho_{x0}$). Also in the computer program, Eq. (12) was used to obtain $R_0 = 1.35 \times 10^{28}$ cm with $(\dot{R}/c)_0 \approx 1$ in good agreement with the geometric consequences. The predicted Hubble constant $H_0 = 68.6$ km s$^{-1}$ Mpc$^{-1}$ is in good agreement with WMAP measurements and the deceleration of the universe $q_0 = 0.005$ is in good agreement with the steady-state prediction of the geometric consequences. From the beginning of the expansion to its steady-state limit, the range of tH is $1/2 \leq tH \leq 1$ and the range of $\Omega_{Tot}$ is $1 \geq \Omega_{Tot} \geq 1/4$.

An expression for the 4-D planckton production rate at any cosmic time t in the expansion is obtained from content consequences by solving for $\dot{N}_4$ from the geometric relation for H in Eq. (5a) and then substituting H from Eq. (10) to get,

$$\dot{N}_{4\,uC} = 4(N_4/t)(\rho_T/\rho_{T2}). \tag{13}$$

To demonstrate consistency, this expression in terms of cosmic time and changing densities, can be shown to be the same as the geometric Eq. (5b), $\dot{N}_{4\,uG} = (N_3/t_p)(\dot{R}/c)$. First, solve Eq. (12a) for $\rho_T/\rho_{T2} = (\dot{R}/c)/(R/ct)$. Then, set Eq. (5a) = Eq. (5b) to obtain $(4N_4)=(N_3/t_p)(R/c)$ and on substitution into Eq. (13), obtain $\dot{N}_{4\,uC} = (N_3/t_p)(\dot{R}/c)$, the same as Eq.(5b).

Thus, in principle, the SC-expansion model is complete with no free parameters. The input constants may need adjustments for early pressure and the crude nucleosynthesis extraction of the present baryon density ($\Omega_{B0}=0.031$) is shy of the WMAP value of 0.04.

**6.2 Fundamental Consequences** Now we come to an observation of the new SC-model that has very important consequences for understanding many puzzling features of our 3-D universe. Immediately after the 4-D ball was formed, $Z \approx 2. \times 10^{26}$, the SC-model predicts the radius of our universe was expanding at the rate $\dot{R} = 1.95 \times 10^{24}c$. The BB-model also predicts very high early expansion rates, but there is a major fundamental difference. The BB-model, with only three spatial dimensions, does not allow even an empty higher dimensional space to support the expansion.

However the SC-model has the pre-existing cellular epi-space into which the 4-D ball is expanding. If our 3-D universe will not allow any energy to move locally faster than the speed of light, then a similar constraint in the epi-universe means its equivalent maximum local speed of epi-communication is $c^+ > 10^{24}$ c. Thus the puzzle of large-scale homogeneity of matter in our universe is readily explained with that high-speed $c^+$ during formation. Note this argument does not necessarily apply to the non-interacting 4-D x-stuff or dark mass [5]. Fundamental consequences of such a present high $c^+$ epi-communication, not yet investigated, include the epi-source of quantum behavior, the epi-source of the constancy of the 3-D velocity of light and other effects of two interacting universes.

The vision of our universe was of cellular spaces and a discrete time, but many of the equations are in terms of rates, a differential term of continuum mathematics. Can an integer equation be derived in terms of the number of cells and the number of ticks of time? Setting $\dot{N}_{4\,uG}$ of Eq. (5b) for the geometric model equal to $\dot{N}_{4\,uC}$ of Eq. (13) for the contents model, for the limit case of $\dot{R}/c = \rho_T/\rho_{T2} = 1$ gives the integer equation,

$$N_{4u} = \tfrac{1}{4} N_{3u}(t/t_p) = \tfrac{1}{4} N_{3u}N_t, \tag{14}$$

where $N_t$ represents the total ticks at time t of a hypothetical Planck clock that made one tick every Planck second ($t_p=0.54 \times 10^{-43}$ s) since the very first 4-D cell was produced. If Eq. (14) is returned to normal units (R, t), it translates to R=ct, a surprising correspondence with relativity. If units of c=1 ly/y, the limiting fourth SC-spatial dimension is R=t, and so it is for the closed-universe solution k=+1 of relativity theory.

SC-values for total mass energy and vacuum energy, and their densities, are plotted in Fig. 1 versus the expansion. Vacuum energy [and time] can be extended back to $N_4=1$.

There are many predictions and consequences of the SC-model that cannot be developed in this paper such as a new "expansion force" [2, 5] and black holes without internal singularities [2, 4]. However, more must be explained about dark mass before we turn to the Supernova Ia and the current accelerated expansion.

**6.3 Dark Mass (DM)** All dark mass is formed into clumps in the last act of creation of the 4-D ball and that initial distribution sets (roughly) the present large-scale structure of galaxies, clusters and voids. Dark mass does not consist of 3-D particles; indeed, it is not even 3-D stuff. As explained, it is a variant 4-D spatial cell that is continually rejected by the 4-D ball. The predicted scaling of dark mass with the expansion of $\rho_x(R)=\rho_{x0}(R_0/R)^2$ and $M_x=M_{x0}/(1+Z)$ with $M_{x0}=6\pi^2\rho_{x0}R_0^3$ and $\dot{M}_x=HM_x$ carries over to the same 3-D distribution function in growing DM-clumps, $\rho_x(r)=\rho_{x0}(r_0/r)^2$, as astronomers found for the outer radii of galaxy haloes. However the cusp of this distribution at r=0, makes these DM-clumps very susceptible to the formation of very early growing black holes in which the black hole may disappear into the 4-D ball and the remaining hollow clump leads to soft-core galaxies.

The SC-model predicts a beginning 4-D ball (R=72 cm) of total mass energy density $\rho_T=1.17 \times 10^{72}$ g cm$^{-3}$, which is mostly radiation with a fractional dark mass density of $\rho_x/\rho_T=6.6 \times 10^{-50}$. Even so, assuming each 4-D cell of dark mass has an effective 3-D Planck mass, this dark mass is sufficient to produce 100 billion black holes of mass $M_{BH}=3M_{sun}$ by $Z=10^{12}$, which could grow to $4.4 \times 10^8 M_{sun}$ by decoupling at Z=1100 and to $5.4 \times 10^{11} M_{sun}$ dark mass black holes at the present. Of course, all of the dark mass does not go to produce, or enlarge, a black hole because much grows outside the black hole to expand the halo of the proto-galaxy. Perhaps there was no early completely "dark age."

These and other considerations (including the rejection of the BB-inflation) lead to the prediction that the $10^{-5}$ level pattern of the CMB will be found to be the initial, as yet unpredictable, distribution of the dark mass clumps where many have already formed dark mass back holes with some Eddington-limited accretion of matter. Also because of the SC-expansion force [2, 5] mentioned earlier, the SC-model predicts that matter ($\Omega_m \sim 0.03$) could not have condensed gravitationally without these dark mass seeds.

**6.4 Supernova Ia (SNIa)** With this context, we come to the key physical phenomena of the exploding stars called "Supernova Ia" [15] to test the predictions of the new spatial condensation model to confirm that no accelerated expansion is necessary.

A radial uniform pulse (few months) of radiation of varying luminosity L leaves the exploding SNIa source as an expanding 2-sphere. The arriving flux of radiation at the detector at nearby radius r is $F=L/(4\pi r^2)$. If the pulse was emitted at time $t_{em}$ and arrived at time $t_0$, then radius $r=c(t_0-t_{em})$. However, if the source was at a great "emission distance,

ED" and redshift Z, then three other effects combine to reduce the energy flux. Let "$d_L$" be the "luminosity distance" adjusted for these three effects.

The expansion degrades the flux of photon energy by a redshift factor of $1/(1+Z)$ and another factor of $1/(1+Z)$ for a time dilation of the time interval between incoming photons. The third effect due to the expansion is calculated differently for the non-relativistic SC-model than from the BB-model [16]. Between emission and detection $(t_0 - t_{em})$, expansion moves the source to a greater "reception distance, RD" where $RD=(1+Z)ED$ [17]. So the effective increase in photon distance due to the expansion is $RD-ED = Z \cdot ED$ and combining all three effects gives [2, 3],

$$d_L = (c(t_0 - t_{em}) + Z \cdot ED)(1+Z). \qquad (15)$$

The distance modulus is [16],

$$m - M = 5\log(d_L/10 \text{ pc}), \qquad (16)$$

where m is the apparent luminosity of the source and M is its absolute luminosity. Distances are usually expressed in units of $c/H_0$ and so the important details of the cosmological model, R(t), appear in the values of $t_0$, $H_0$ and ED(Z).

Two methods were used to calculate values for $d_L$ and luminosity. Time is defined explicitly in the rather simple SC-model, so to obtain the emission distance as a function of Z, a larger value $Z_{max}$ was selected with an estimated value of $ED(Z_{max})$ and then a trial-and-error integration [3] of the photon trajectory to the detector was repeated [new $ED(Z_{max})$] until the photon arrived at the detector with ED=0 at $t=t_0$ and Z=0. Of course, on that final trajectory, values of the desired parameters were valid for other photons emitted at that Z which also arrived at the same time $t_0$ at the detector.

For the second method, an analytic solution for ED(Z) was obtained for the case where the radiation density could be neglected (Z≤6). This computer program eliminated the laborious trial-and-error calculation with which it agreed.

With $t_0$=13.5 Gy, $H_0$=68.6 km s$^{-1}$ Mpc$^{-1}$, $\Omega_{B0}$=0.031 and $\Omega_{DM0}$=0.248, the good fit of the SC-predicted curve to the 1995 supernova Ia data of Hamuy [18] and to the 1998 Supernova Cosmology Project [19] are shown in Fig 2 and to the High Z Search Team data [20] in Fig. 3. Thus the main goal is accomplished that, with the new spatial condensation model, no accelerated expansion is necessary to explain the data [2, 3].

**6.5 Comparison of Model Parameters** There is a great body of experimental facts that the new theory must also predict correctly and there are many measurements that agree with relativity theory. This section will compare the new SC-model and BB-model predictions of the evolution of the cosmological parameters with special emphasis on the range of evolution important to the SNIa measurements between redshift Z=0 to Z=2. Also any difference that the two models predict for the future will be discussed.

For the SC-model the Hubble parameter, H, is given by Eq. (11) or in terms of redshift Z by Eq. (12b) if it is multiplied by (c/R). The gravitational constant G does not appear and densities of the contents appear as ratios.

For the BB-model, expressions for the Hubble parameter and $\dot{R}/c$ can be derived from the Friedmann equation $\dot{R}^2 = (8\pi G/3)\rho R^2 - kc^2$. Here it is gravity acting on the contents that directly controls the changing scale factor in contrast to the SC-model where the contents, ρ, only moderate the driving vacuum energy. For the BB-model the Hubble parameter is [21],

$$[H(a)]^2 = H_0^2[\Omega_\Lambda + \Omega_m a^{-3} + \Omega_r a^{-4} - (\Omega - 1)a^{-2}], \qquad (17)$$

where a is the ratio of scale factors, $a=R/R_0=1/(1+Z)$ and the $\Omega_i$ are the current densities $\rho_i$ divided by the BB-critical density $\rho_c$, i.e., $\Omega_i = \rho_i/\rho_c$, where $\rho_c = 3H^2/8\pi G$. The subscript lambda, $\Lambda$, implies the constant vacuum energy term that Einstein added to his field equations before he was informed that our universe was really expanding. In terms of redshift Z, Eq. (17) states that,

$$H(Z) = H_0(1+Z)[1 + \Omega_m Z + \Omega_r((1+Z)^2-1) + \Omega_\Lambda((1+Z)^{-2} - 1)]^{1/2}, \qquad (18)$$

In contrast to $\Omega=0.28$ for the SC-model closed universe, most BB-cosmologists [19] insist that our 3-D universe has $\Omega=\Sigma\Omega_i=1$ for an infinite k=0 universe. The SC-model predicts that the future limit of the expansion rate of our universe, given by Eq. (12), is equal to c. For the BB-model, an expression for $\dot{R}/c$ can be derived from Eq. (18) since $R=R_0/(1+Z)$,

$$\dot{R}/c = (R_0 H_0/c)[1 + \Omega_m Z + \Omega_r((1+Z)^2-1) + \Omega_\Lambda((1+Z)^{-2} - 1)]^{1/2}. \qquad (19)$$

For the comparison here, both models will use the same current values of the parameters: mass density, $\Omega_m=0.28$; $H_0=68.6$ (km/s)/Mpc; and $\rho_{c0}=8.85\times10^{-30}$ g cm$^{-3}$. For the BB-model both the baryons and "dark matter" scale as matter, $\propto R^{-3}$, but in the SC-model only the baryons, $\Omega_B=0.03$, scale as matter and the balance "dark mass", $\Omega_x=0.25$, scales with the expansion as $R^{-2}$. To obtain $\Omega=1$, and fit the SNIa data, the BB-cosmologists add $\Omega_\Lambda=0.72$, the energy of Einstein's "lambda" constant.

Our 3-D universe of the SC-model was formed at R~72 cm or $Z=1.88\times10^{26}$. To compare the two models back to that high Z, the radiation term, $\Omega_{r0} = 1.06\times10^{-4}$, must be included. Logarithms to base 10 are used to graphically display Z, H and $\dot{R}/c$. The results in Fig. 4 show that both models predict the same H and $\dot{R}/c$ after radiation becomes dominant at $\log(1+Z) \sim 3$. Of course the BB-curves, as R→0, are predicted to go to ever-greater Z→∞ for a universe with k=0. The interesting region for Z<6 is shown in Fig. 5 versus Z with labels "($\Omega_m, \Omega_\Lambda$)".

In Fig. 5 additional curves are also shown for the BB-model for matter mass alone, BB-(0.28, 0) to show the change gained by the addition of lambda in curve BB-(0.28, 0.72). In the region of Z<3, the added lambda term lowers the BB-H-curve to fair agreement with the SC-H curve which does fit the data as shown in Figs. 2 and 3. The entire future of our 3-D universe is contained between Z=0 and Z=-1, since $R/R_0=1/(1+(-1))= +\infty$.

It is the upper set of BB-curves for the predicted expansion rate that are troubling. At the present, both models agree that $(\dot{R}/c)_0=1$ and acceleration of the BB-expansion is predicted to begin at Z~0.6. It is the BB-predicted future (Z<0) rapid acceleration of the expansion that is ominous and unconvincing while the SC-model needs no dark energy or accelerated expansion. In contrast, the SC-$\dot{R}/c$ curve goes smoothly to a future value of unity (deceleration q=0) for a reasonable steady-state expansion rate.

Although not shown, similar curves for the omegas are consistent, respectively, with the data of Figs. 4 and 5. Both models predict $\Omega=1$ during radiation domination. In contrast for the future, BB-$\Omega=1$, but SC-$\Omega\rightarrow1/4$.

With the added lambda energy of $\Omega_\Lambda=0.72$, the BB-(0.28, 0.72) curve for H of Fig. 5 shows that the two models will be in fair agreement until Z~3 and then at higher Z the predictions diverge. Unfortunately, astronomers expect to find in the near future few SNIa for Z greater than 2. However, as we will see in the next section 6.6, we can still test model predictions against already measured SNIa radiation at much higher Z.

**6.6 Future Measurements of Past Supernova Ia at High Z** This paper will end with an interesting thought experiment for which, amazingly, we already have data to test model predictions at very high Z (>2). The radiation of past supernova Ia, that our astronomers have measured, is still traversing through the universe. In principle, astronomers of the future could measure that same radiation on other planets. That radiation since emission at scale factor $R_{em}$ continues to increase redshift $Z_{em}$ with continued expansion of scale factor $R_0$, $1/(1+Z_{em}) = R_{em}/R_0$ [or decrease in $(R_{em}/R_0)$ if $R_0\equiv1$] .

If the scale factor R of our universe is increased by factor f, the currently measured supernova radiation of redshift Z will then have redshift Z'= f(1+Z) –1 independent of the model. On the other hand, the predicted new values of apparent luminosity $m_{eff}$ and luminosity distance $d_L$ will depend upon the model.

Any cosmological model that predicts the scale factor into the past R(t), can also calculate R(t) into the future and therefore predict, at factor f, values of $t_0$', $H_0$',etc. and therefore, new values $d_L'(Z')$, and $m_{eff}'(Z')$ for the very same radiation of known absolute luminosity M at the higher Z'.

If one has correctly modeled, as a function of Z, all of the factors that decrease the energy flux of the radiation as its distance increases from the source, then the expression for the luminosity distance $dL/(c/H_0)$ as a function of Z, assuming no energetic collisions, should fall on a universal curve. Also, any one specific spherical expanding pulse of SNIa radiation should slowly move up the universal curve with time.

This exercise was easily carried out with the SC-model and the results are shown in Fig. 6. The exercise could have been demonstrated, in principle, by just one supernova Ia, but a set of real data was considered more effective. The early 1995, 18-points, low Z, measurements of Hamuy, et al, [see Fig. 2] were selected and a least squares [$m_{eff}$ vs log Z] fit used to obtain residual values for the $m_{eff}$ data points. For four different f-values (1, 2, 3, 6), the SC-computer program predicted SC-curves for $d_L/(c/H_0)$ and $m_{eff}$ and predicted values for the data points. Assuming future astronomers with equal accuracy of measurement, the residuals were added back for visual effect.

As shown in Fig. 6, the finding for the SC-model, with its steady-state expansion rate, is that the luminosity distance $d_L/(c/H_0)$ vs Z is indeed a universal curve where the data point for the expanding pulse of radiation, from any one supernova Ia source, just moves up this curve as the universe expands. Of course, future measured values of flux decrease, (effective magnitude $m_{eff}$' increases) of the same radiation. The symbols show the predicted future $m_{eff}$' of the radiation of the 18 Hamuy supernova. Only two predicted $m_{eff}$' curves are shown for f=1 and f=6 [limit of ancient graphics software]. The f=1, $m_{eff}$ curve of Fig 6 and Hamuy data points are the same as in Fig. 2. It would be interesting for comparison, to see what the big bang theoreticians with their dark energy predict for such future measurements.

Figure 6 predicts measurements of existing SNIa radiation for future astronomers. One could also ask what the SC-model predicts for existing SNIa radiation by possible astronomers of the past. For this exercise, seven "gold, high Z" (1.140 to 1.551) SNIa measurements [22] were selected and the same computer program was used to calculate predicted measurements for $f=R/R_0=0.5$ at $t=6.402$ Gy. Both present and past predicted values of $m_{eff}$ and $dL/(c/H_0)$ are shown in Fig. 7 where present values of luminosity $m_{eff} = (m-M) +(-19.34)$. Again the present $m_{eff}$ curve is the same as that of Fig. 2 for which the present data fit well and the present values of $dL/(c/H_0)$ fall on the universal curve as do the projected past measurements at $f=R/R_0=0.5$. The predicted past values of $m_{eff}$ are smaller showing increased flux closer to the source. As a final test run, f was decreased to 0.40 ($t=5.0$ Gy) corresponding to $Z=1.5$ and, indeed, only the highest SNIa of $Z=1.551$ survived the calculation (see circled symbol near the origin). The six other lower-Z SNIa were rejected because at $Z=1.5$ the "past astronomers" were in our universe before the six lower-Z SNIa occurred.

## 7. Concluding Remarks

The new concept of spatial condensation of a substantial space was the key that opened the door to the internal machinery that started, and now drives, the expansion of our universe. The geometric calculation supported the assumption of a closed geometry but it was the SC-contents model that finally exposed all of the beautifully synchronized motions of its internal parts. An added fourth spatial dimension molds the important curvature of our 3-D space and a new scaling factor for dark mass set a reasonable limit for the expansion rate. Dark energy is not needed, instead, the BB-prediction of accelerated expansion signals the first clear global failure of relativity theory.

The new definitions and understanding of the fundamental physical concepts of space, time, energy, mass, inertia and the source of the attractive attribute of gravity, has given this effort its own reward. With no need for accelerated expansion, there is much more to learn from the SC-model. The realization that we exist simultaneously in two very different interacting universes opens a new theoretical window. However, unraveling the epi-physics of the epi-universe could take decades. Understanding the details of spatial condensation and quantum behavior will be rewarding, but with little hope of direct penetration into the different spaces on either side of our 3-D universe, that task will be difficult.

## 8. Acknowledgments

The author thanks his good friend, Emeritus Professor Robert A. Piccirelli, for extensive discussions of the new physical concepts.

**LEGENDS**

**Fig. 1** The spatial condensation (SC) model predicts that dark mass (not matter) of the universe increases and overtakes matter mass at size $R/R_0$ ~10 (Z~9) but the total mass density continues to decrease with the expansion. The predicted vacuum energy (no mass) which drives the expansion (no acceleration) continues to increase and its future density approaches a fundamental geometric constant value $e_v = \hbar/l_p^3 t_p = 4.635 \times 10^{114}$ ergs/cm$^3$.

**Fig. 2** The predicted apparent magnitude $m_{eff}$ curve of the SC-model [$\Omega_{Tot} = \Omega_{mass} = 0.28$, $t_0 = 13.5$ Gy, h=0.686 [3] is compared to the early 1995 low-Z supernova data of Hamuy [18] and the higher-Z, 1998 supernova data of the Supernova Cosmological Project (for fit: $\Omega_{Tot} = 1$, $\Omega_m \approx 0.28$, $t_0 \approx 14.9$, h=0.63, [19]). This predicted SC-curve shows that no dark energy or accelerated expansion is needed to fit the SNIa data.

**Fig. 3** The predicted distance modulus m-M curve of the SC-model [(M=-19.34 and $\Omega$, $t_0$, h of Fig. 2] is compared to the High Z Search Team 1998 data [20] and later data (for fit: $\Omega_{Tot} \approx 1$, $\Omega_m \approx 0.29$, [22]). This predicted SC-curve also shows that no dark energy or accelerated expansion is needed to fit the SNIa data.

**Fig. 4** The accuracy of a cosmological model is contained in its scale factor function R(t) or Hubble parameter H= $\dot{R}$/R. This figure shows that the SC-model and the big bang (BB) model agree during the early universe of radiation domination (Z>3000). The interesting low-Z period of matter and dark mass domination is shown in greater detail in next Fig. 5.

**Fig. 5** Significant differences between the SC-model and BB-model are shown for expansion rate expressed as $\dot{R}$/c and the Hubble parameter. Labels for curves are expressed in terms of two dimensionless densities ($\Omega_{mass}$, $\Omega_\Lambda$) where $\Omega_{mass}$ includes baryon mass and dark mass and $\Omega_\Lambda$ is the additional "dark energy" required by the BB-model to fit the supernova Ia data. The BB-model without the added $\Omega_\Lambda$-energy is labeled ($\Omega_{mass}$, 0) and did not fit the data. The SC-(0.28, 0) curves, used in Figs. 2 and 3 are well behaved into the future (Z=0 to –1) but the BB-(0.28, 0.72) curve, adjusted to fit the data, predicts trouble into the future.

**Fig. 6** Supernova Ia data at higher Z await better instruments but consistency of model predictions at higher Z can be obtained with existing data projected for future astronomers on distant planets. Consider the expanding 2-spheres of the low-Z, SNIa radiation measured by Hamuy, et al. of Fig. 2. If the scale factor R of our universe increases by a factor f, then the new Z` of that radiation is Z`=f(1-Z)-1. The SC-model predicts the luminosity distance just moves up the $d_L/(c/H_0)$ universal curve while the effective luminosity $m_{eff}$ increases with decreasing flux. What does the present $\Lambda$CDM BB-model predict at these higher Z?

**Fig. 7** With the same computer program used for Fig. 6, seven of the "Gold, High-Z" recent SNIa measurements of Riess, et al. [22] were predicted for past astronomer measurements at f=0.50 and t = 6.42 Gy. Again the same universal curve for $d_L/(c/H_0)$ is confirmed with $m_{eff}$ lower (energy flux higher) closer to the source. With f=0.40, and t=5.0 Gy, only the highest Z=1.551 SNIa appears because the other six have not yet exploded.

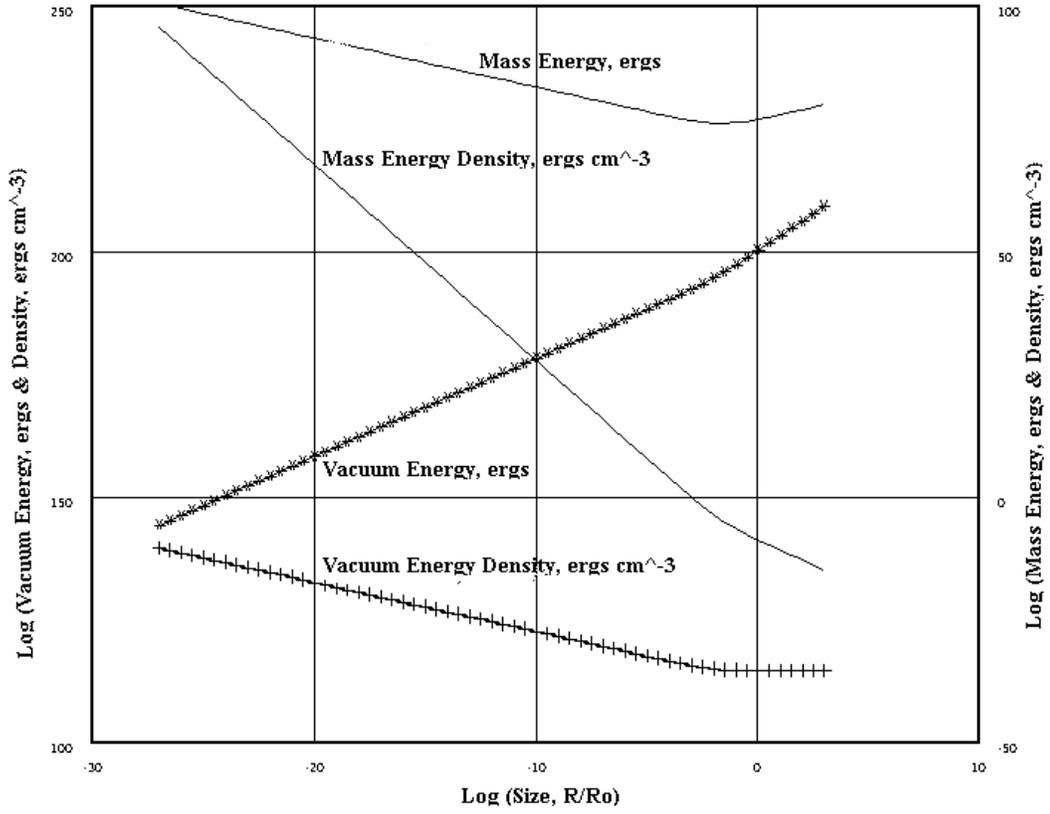

Figure 1

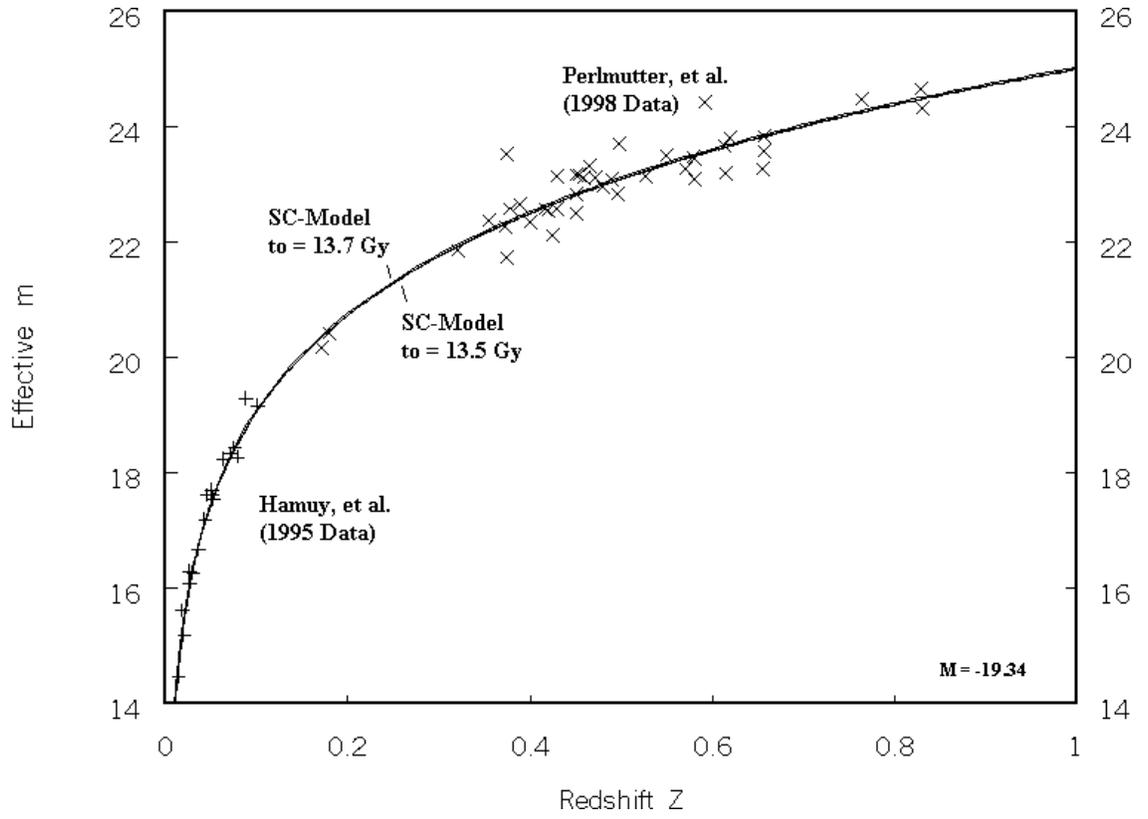

Figure 2

x

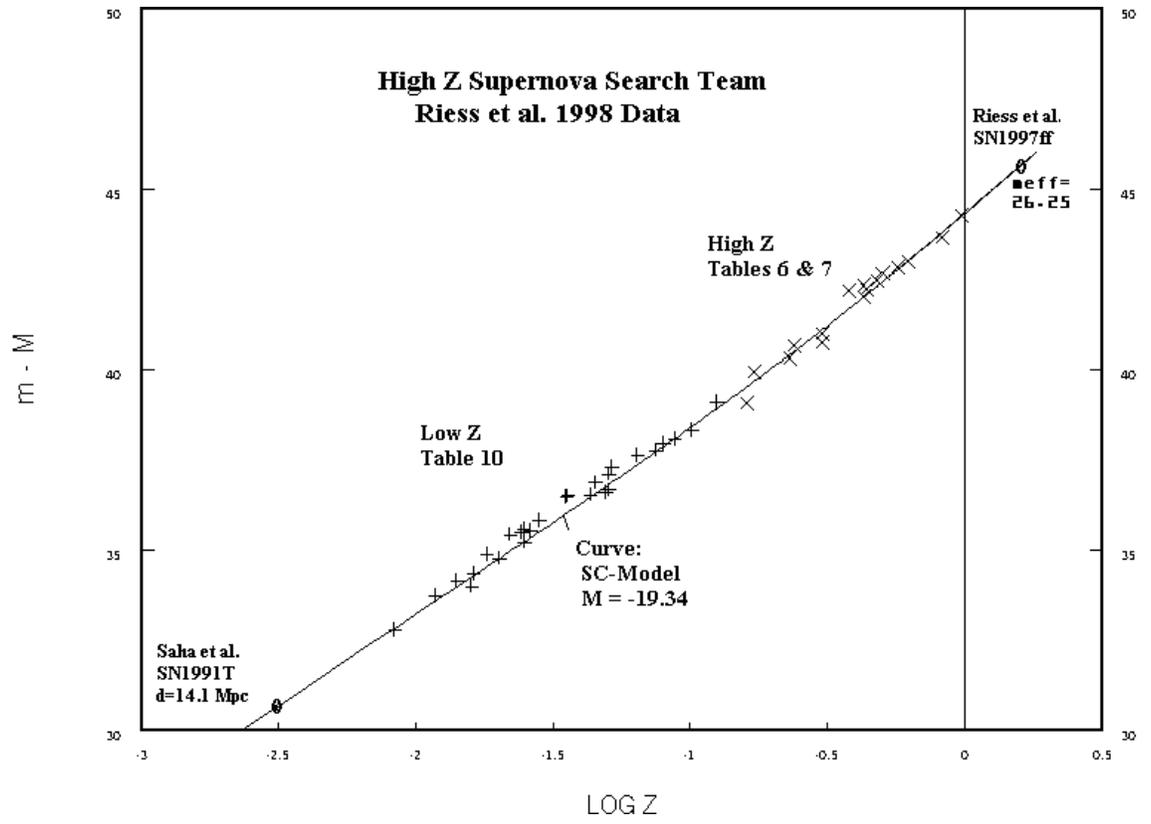

Figure 3

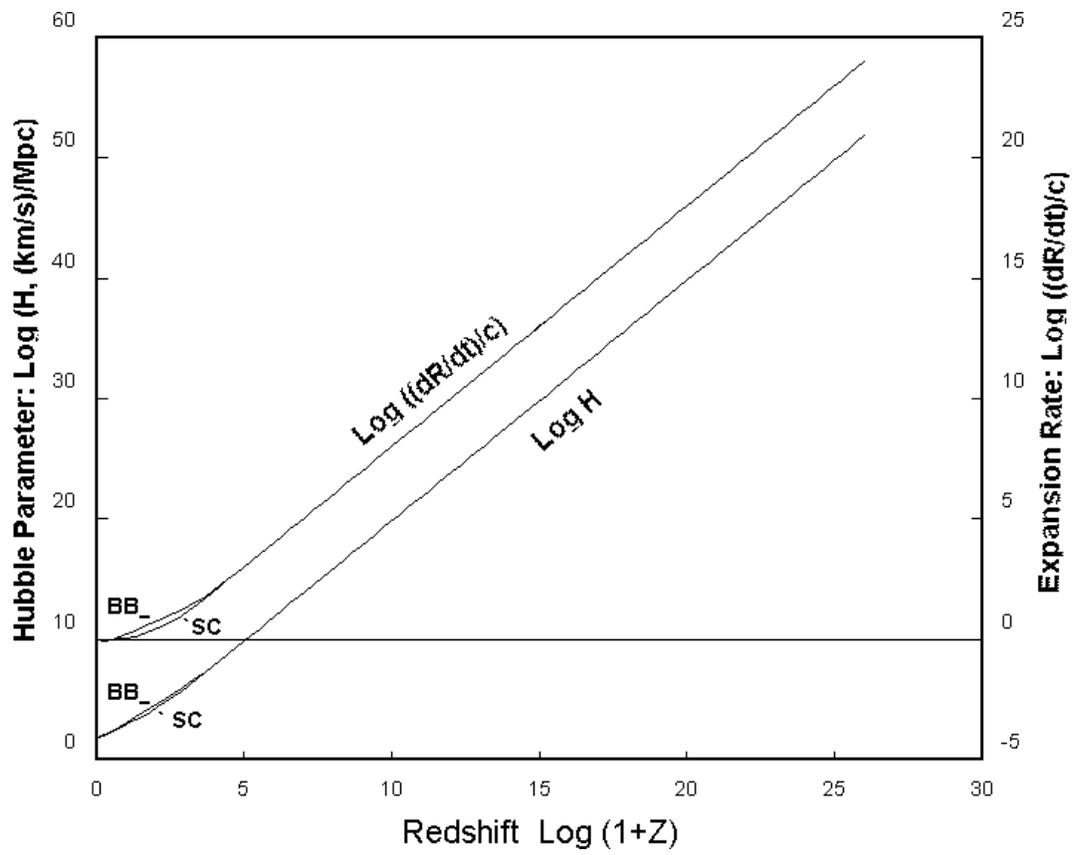

Figure 4

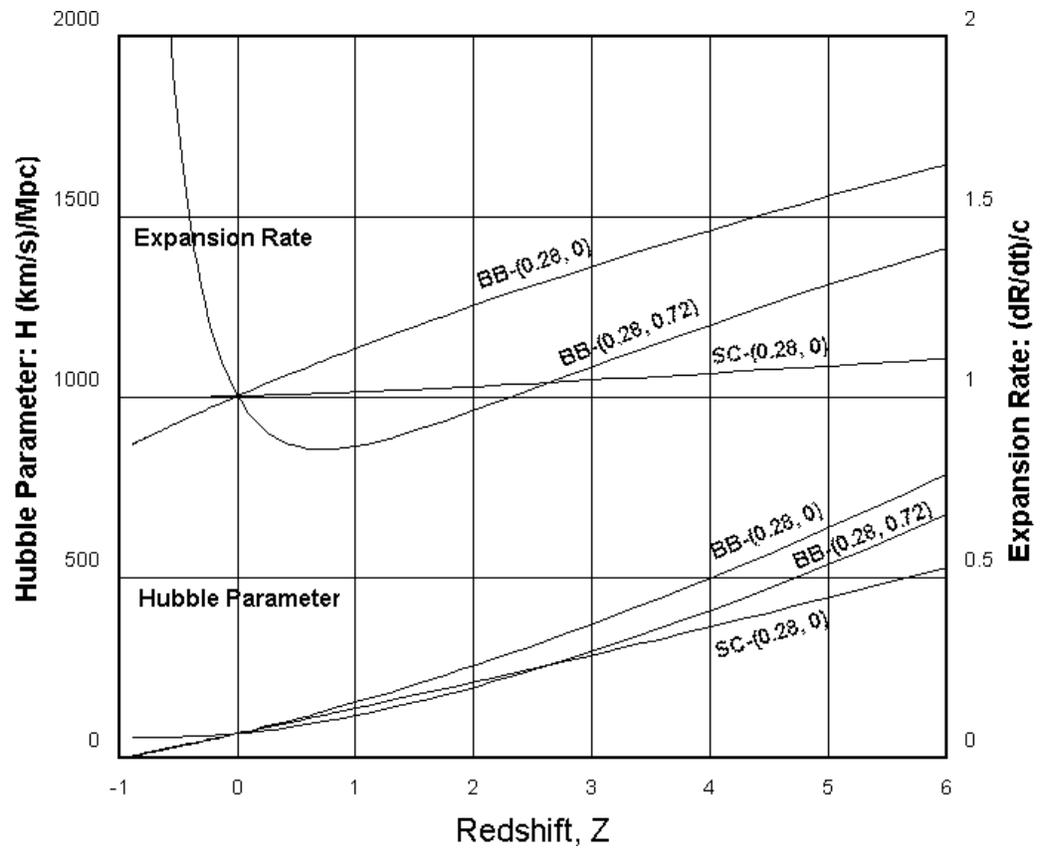

Figure 5

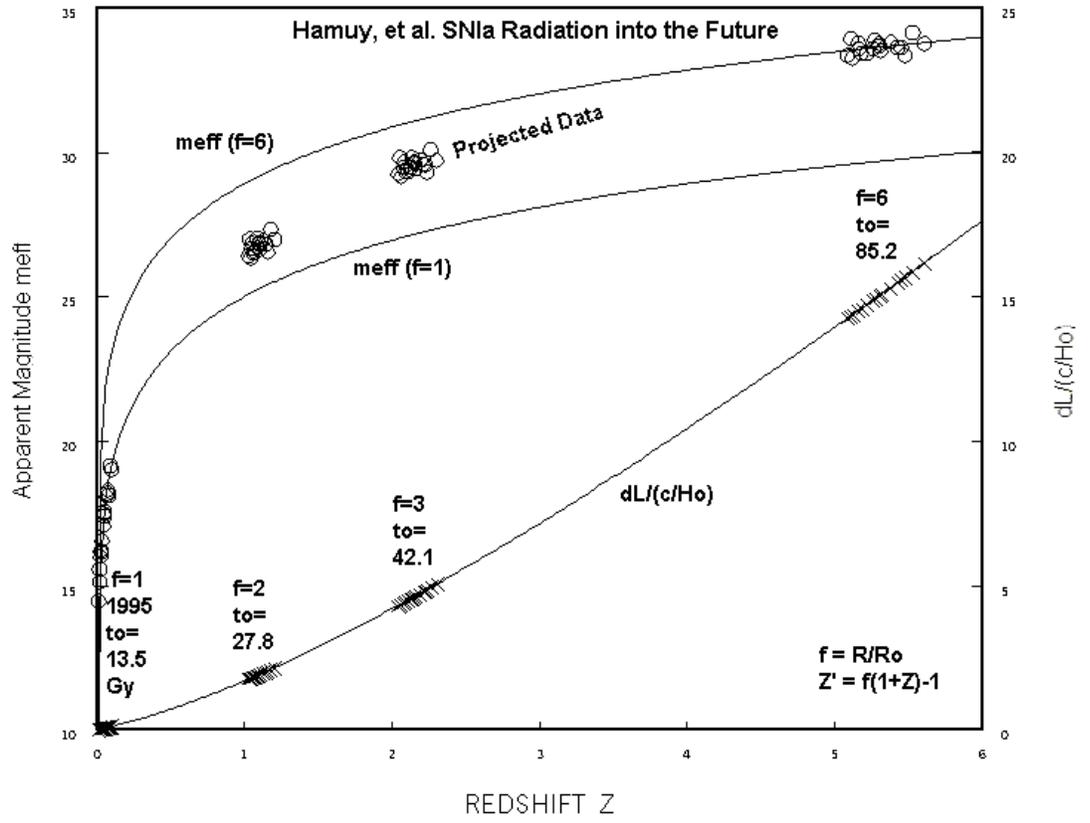

Figure 6

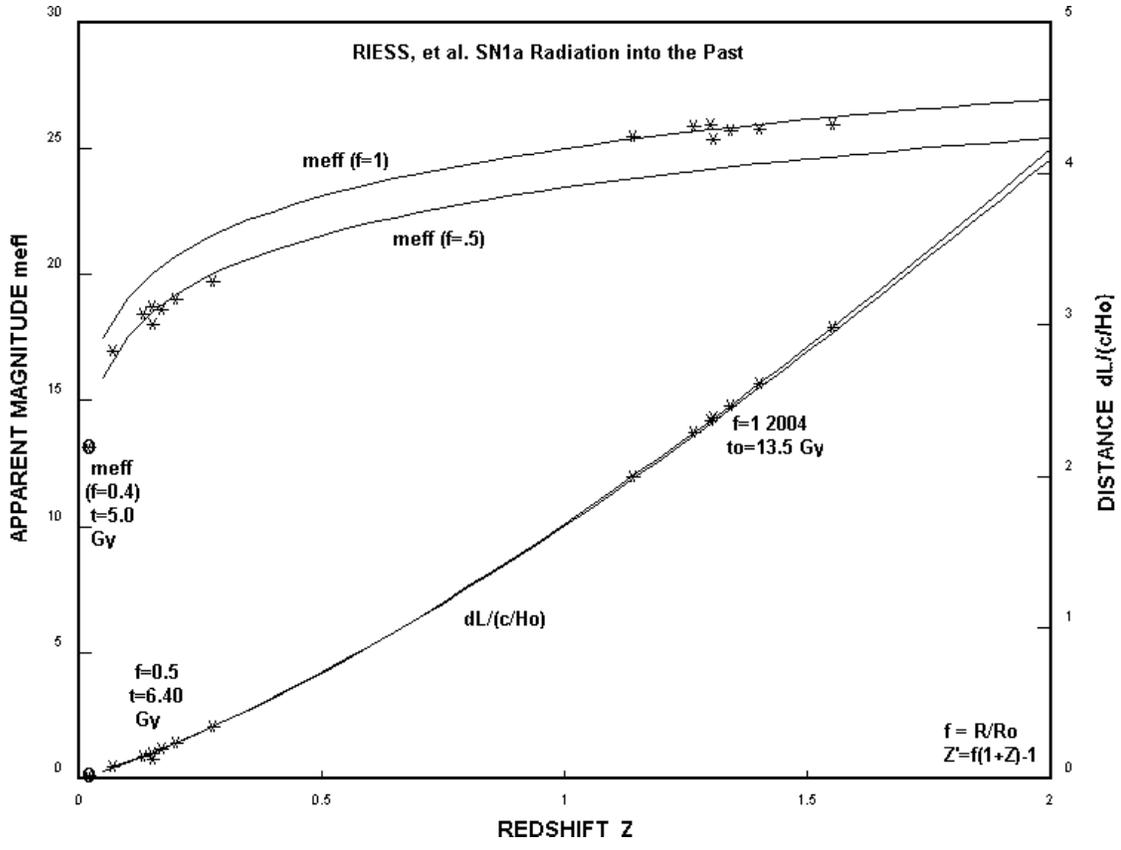

Figure 7